\documentclass[lettersize,journal]{IEEEtran}
  \usepackage{amsmath,amsfonts}

  \usepackage{array}
  \usepackage{textcomp}
  \usepackage{stfloats}
  \usepackage{url}
  \usepackage{verbatim}

  \usepackage{graphicx}
  \usepackage{subcaption}
  \usepackage{cite}
  \usepackage{stmaryrd}
  \usepackage{threeparttable}
\usepackage[ruled,vlined,linesnumbered]{algorithm2e}
\SetKwInput{KwIn}{Input}
\SetKwInput{KwOut}{Output}
\usepackage{circledsteps}

  \usepackage{lipsum}
  \usepackage{textcomp}
  \usepackage{pifont}
  \usepackage{multirow}
  \usepackage{amssymb}
  \usepackage{booktabs}
  \usepackage{siunitx}
  \usepackage{circledsteps}
  \usepackage{orcidlink}
  \usepackage{hyperref}
  \hypersetup{
      colorlinks=true,
      linkcolor=blue,
      citecolor=blue,
      urlcolor=blue
  }

  \usepackage{xcolor}
  \usepackage{empheq}

  \newcommand{\rom}[1]{\uppercase\expandafter{\romannumeral #1\relax}}

\begin{document}

  \title{An Efficient Fault-Tolerance Scheme for \\CKKS Computation on CPUs}

  \author{Jianan Mu$^{\orcidlink{0000-0001-8513-0792}}$,
  Ge Yu$^{\orcidlink{0009-0008-3556-3774}}$,
  Tenghui Hua$^{\orcidlink{0009-0008-3476-9367}}$,
  Liang Kong$^{\orcidlink{0009-0009-6031-4312}}$,
  Jing Ye$^{\orcidlink{0000-0002-8023-5090}}$,
  Xing Hu,
  Meng Li,\\
  Xiaowei Li$^{\orcidlink{0000-0002-0874-814X}}$,~\IEEEmembership{Senior Member,~IEEE},
  and Huawei Li$^{\orcidlink{0000-0001-8082-4218}}$,~\IEEEmembership{Senior Member,~IEEE}
  \thanks{Jianan Mu, Ge Yu, Tenghui Hua,  Jing Ye, Xing Hu, Xiaowei Li, Huawei Li (State Key Laboratory of Processors) are with the Institute of Computing Technology, Chinese Academy of Sciences, Beijing
  100190, China, and also with the University of Chinese Academy of Sciences, Beijing 100190, China. Liang Kong is with the Tsinghua University, Beijing 100084, China. Meng Li is with the Peking University, Beijing 100871, China.}%
  }


  \markboth{IEEE TRANSACTIONS ON COMPUTER-AIDED DESIGN OF INTEGRATED CIRCUITS AND SYSTEMS}%
  {Shell \MakeLowercase{\textit{et al.}}: A Sample Article Using IEEEtran.cls for IEEE Journals}


  \maketitle
  \pagestyle{plain}

  \begin{abstract}
Fully homomorphic encryption (FHE) enables computation directly
on encrypted data, but its long ciphertext dataflow and
high-dimensional modular arithmetic make it highly vulnerable to
silent data corruption (SDC) caused by transient hardware faults.
Although recent fault-tolerance schemes have been explored for
dedicated FHE accelerators, efficient reliability protection for
CPU-based FHE software remains underexplored. Direct redundant
execution more than doubles the runtime, while directly applying
existing checksum-based protection to CPUs still incurs substantial
modular-arithmetic and memory-access overheads.

This work presents an efficient fault-tolerance scheme
specifically designed for CKKS computation on general-purpose
CPUs.
The proposed scheme protects CKKS polynomial operators by checking their
input--output consistency and reduces the associated overhead at three
levels.
First, modulus-aware bucket checksum exploits wide CPU accumulators to
reduce expensive modular reductions.
Second, dataflow-fused in-operator checking integrates checksum accumulation
into the original operator dataflow, eliminating detached scans of long
ciphertext polynomials.
Third, cross-operator check fusion removes redundant checksum computation
across adjacent polynomial operators while preserving the end-to-end
checking invariant.

We implement the proposed scheme in OpenFHE and evaluate it
on representative encrypted applications and ciphertext primitives
under random single-bit transient faults. The proposed scheme
achieves a 100\% empirical detection rate over 150,000 evaluated
non-crashing corrupted-result cases and maintains application
accuracy close to the fault-free baseline across a wide range of
fault rates. Meanwhile, it incurs only 6.0\%--8.4\% runtime
overhead, averaging 6.8\%, and reduces the average protection
overhead by 4.9$\times$ compared with direct checksum-based
protection.
\end{abstract}
  \section{Introduction}
\label{sec:intro}

\begin{figure}[t]
    \centering
    \includegraphics[width=1.0\linewidth]{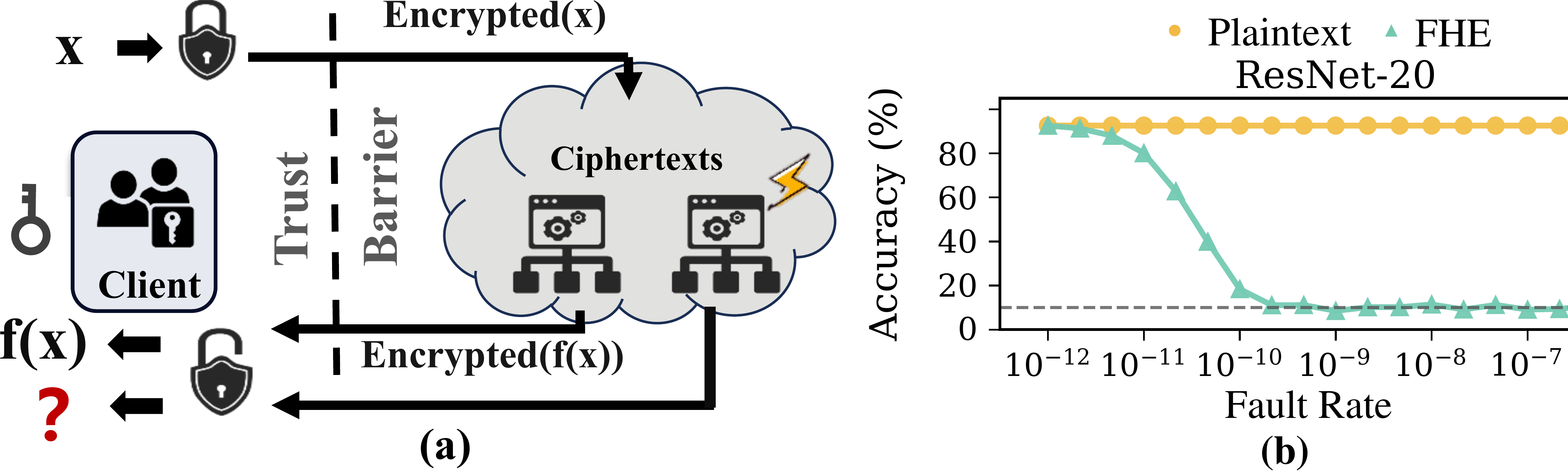}
    \caption{(a) Typical application scenarios of FHE. (b) Fault injection evaluation on CKKS-encrypted ResNet-20.}
    \label{fig:fhe_app}
    \vspace{-0cm}
\end{figure}

Fully Homomorphic Encryption (FHE)~\cite{FHE,BFV,BGV,TFHE,CKKS} is an emerging cryptographic technique that enables computation directly on encrypted data.
It allows clients to outsource computation to untrusted servers while keeping data encrypted throughout storage, transmission, and processing, as shown in Fig.~\ref{fig:fhe_app}(a).
This property makes FHE a promising foundation for privacy-preserving services such as encrypted finance, biomedical analytics, private databases, and secure artificial intelligence (AI) inference~\cite{FHEFinancial,FHEMedical,HE3DB,FHEML,CKKSCNN,HELR,LoLa}.
The importance of FHE has also been recognized by the broader systems community: the SPEC CPU 2026 benchmark suite includes a homomorphic-encryption benchmark, \texttt{750.sealcrypto\_r}, based on encrypted database queries~\cite{SPEC_CPU2026}.
Among existing FHE schemes, the Cheon-Kim-Kim-Song (CKKS) scheme~\cite{CKKS} is particularly important because it supports single-instruction multiple-data (SIMD)-style approximate arithmetic, making it suitable for numerical and AI workloads.
While recent FHE research has primarily focused on improving execution performance, the reliability of FHE algorithms on real-world hardware remains insufficiently studied.

Real-world hardware inevitably experiences transient faults that may cause
\textit{silent data corruption (SDC)}: erroneous outputs without crashes or visible error reports~\cite{papadimitriou2023silent,dixit2021silent}.
Such faults are particularly concerning for FHE computation.
On the one hand, FHE programs execute long chains of high-dimensional modular polynomial operations, making their execution much longer than plaintext computation and increasing the exposure window to transient faults.
On the other hand, during outsourced FHE execution, the server cannot inspect plaintext semantics: ciphertexts appear pseudorandom, and a corrupted ciphertext may remain syntactically valid until it is decrypted by the client.
As shown in Fig.~\ref{fig:fhe_app}(b), our preliminary fault-injection experiment on ResNet-20 shows that, for the same inference task, the CKKS-encrypted execution is substantially more vulnerable to hardware faults than the plaintext execution.
While the plaintext model only exhibits a mild accuracy drop under fault injection, the encrypted model suffers a sharp accuracy degradation after decryption.
These observations motivate the need for efficient SDC detection in FHE computation.

Recent security and architecture studies have begun to investigate this reliability problem, especially for FHE accelerators, simulator-based accelerator designs, and ciphertext storage under memory faults~\cite{glitchfhe,reliafhe,reliaanalysis}.
However, CPU-based FHE software remains a critical but underexplored target for reliability protection.
At present, CPUs are still widely used in FHE development, testing, algorithm prototyping, and lightweight encrypted-service deployment.
Representative FHE software ecosystems, including Microsoft Simple Encrypted Arithmetic Library (SEAL)-based workloads in SPEC CPU 2026, OpenFHE, and Zama's Torus Fully Homomorphic Encryption (TFHE)/Concrete software stack, all provide CPU-based implementations as part of their software backends~\cite{SPEC_CPU2026,openfhe,TFHE-rs,ConcreteML}.
Even as GPU libraries and dedicated FHE accelerators continue to mature~\cite{F1,ARK,Sharp,CraterLake,MAD,BitPacker}, CPU execution remains an important baseline for practical FHE software stacks.
Therefore, protecting CPU-based FHE computation is necessary for dependable computation on ciphertext.

The challenge, however, is not whether FHE computation can be protected, but whether it can be protected efficiently on CPUs.
Straightforward dual modular redundancy (DMR) can detect computation faults by duplicating the execution, but it approximately doubles the runtime.
Lightweight algorithm-level error-detection methods can reduce the amount of redundant computation by checking input-output consistency of numerical kernels~\cite{ABFT,wang1994FFT,abdelmonem2025efficient}, and recent FHE accelerator protection schemes further exploit customized hardware support~\cite{reliafhe}.
Nevertheless, as quantified in Sec.~\ref{sec:challenge-cpu-overhead}, directly applying these protection mechanisms to CPU-based FHE software still introduces substantial overhead, around 30\%--40\% in our evaluation.
This shows that efficient CPU-based FHE protection cannot be obtained by simply reusing existing approaches.

The root cause is that low algorithmic complexity does not necessarily translate into low CPU execution overhead.
Application-specific integrated circuit (ASIC) accelerators can provide customized parallelism, dedicated modular arithmetic units, and specialized datapaths for protection logic~\cite{F1,ARK,Sharp,CraterLake,MAD,BitPacker}.
CPUs, in contrast, must execute the additional detection logic through fixed instruction pipelines, limited integer arithmetic resources, and cache-based memory hierarchies.
As a result, a protection scheme that is efficient on an accelerator may still incur many extra instructions, frequent modular reductions, and cache-unfriendly memory scans on CPUs.
This motivates an efficient protection design for CKKS software
execution on CPUs: beyond reducing the amount of redundant computation, the detection logic must be reshaped according to CPU resource constraints and the original CKKS operator flow.

To address this problem, we design and implement an efficient
fault-tolerance scheme for CKKS-based FHE programs running
on CPUs.
To the best of our knowledge, this is the first work to
systematically design and optimize fault-tolerance protection
for CPU-based CKKS computation.
Rather than directly porting accelerator-oriented protection
mechanisms, the proposed scheme restructures checksum
computation according to CPU arithmetic resources,
memory-access behavior, and the operator dataflow of CKKS
computation.
Our evaluation shows that the resulting scheme incurs only
6.0\%--8.4\% runtime overhead, averaging 6.8\%, and reduces the
average protection overhead by 4.9$\times$ compared with direct
checksum-based protection.
This paper makes the following contributions:
\begin{itemize}
    \item \textbf{Fault-injection characterization of CPU-based CKKS.}
    We conduct CPU-based fault-injection experiments showing that CKKS computation is highly vulnerable to transient faults, and that single-bit errors can silently cause large application-level deviations.

\item \textbf{Efficient fault-tolerance scheme for CPU-based CKKS.}
To the best of our knowledge, we present the first
fault-tolerance scheme specifically designed to provide
low-overhead protection for CKKS computation on
general-purpose CPUs.
    We identify three major sources of protection overhead---frequent
    modular reductions, detached polynomial scans, and redundant
    operator-boundary checks---and address them through modulus-aware
    bucket checksum, dataflow-fused in-operator checking, and
    cross-operator check fusion, respectively.

    \item \textbf{OpenFHE implementation and evaluation.}
    We implement the proposed techniques in OpenFHE~\cite{openfhe} and evaluate them on representative CKKS applications and primitives.
The results show that the proposed scheme reduces the average
protection overhead from 33.4\% for direct checksum-based
protection to 6.8\%, while significantly improving reliability.
\end{itemize}

  \section{Background and related Work}
\label{sec:preliminary}

\subsection{CKKS scheme and computation}
\label{sec:pre_ckks}

CKKS scheme is an approximate-number FHE scheme that supports SIMD-style arithmetic over encrypted vectors~\cite{CKKS}.
A plaintext vector is first encoded into a polynomial and then encrypted into a Ring Learning With Errors (RLWE) ciphertext.
Homomorphic evaluation is performed through ciphertext-level primitives, which are further decomposed into modular polynomial kernels~\cite{CKKS,RNS-CKKS}.

\subsubsection{Ciphertext data structure}

Let $\mathcal{R}_Q=\mathbb{Z}_Q[X]/(X^N+1)$ denote the polynomial ring, where $N$ is the polynomial degree and $Q$ is the ciphertext modulus.
A CKKS ciphertext is usually represented as $\mathbf{ct}=(c_0,c_1)\in\mathcal{R}_Q^2$, where each component is an $N$-coefficient polynomial~\cite{CKKS}.

Practical CKKS libraries adopt the Residue Number System (RNS) to decompose the large modulus $Q$ into several machine-word-sized primes, i.e., $Q=\prod_{i=0}^{\ell}q_i$~\cite{RNS-CKKS}.
Thus, each polynomial is stored as multiple residue towers, denoted as $c(X)\Rightarrow\{c^{(0)}(X),c^{(1)}(X),\ldots,c^{(\ell)}(X)\}$, where $c^{(i)}(X)=c(X)\bmod q_i$.
This representation enables efficient modular arithmetic over small primes, but it also increases the amount of data that must be accessed and protected.
In this work, CKKS data are viewed at three levels: ciphertext components, RNS polynomial towers, and modular coefficients.

\subsubsection{Polynomial operators}

CKKS computation is dominated by several polynomial-level operators, including the Number Theoretic Transform and its inverse transform, basis conversion, and element-wise multiplication~\cite{CKKS,RNS-CKKS,Bootstrapping}.
We refer to them as Number Theoretic Transform/inverse Number Theoretic Transform (NTT/INTT), basis conversion (BConv), and element-wise multiplication (EWM), respectively.
These operators are the major computation kernels in ciphertext multiplication, rotation, key switching, and bootstrapping.

\textbf{NTT.}
The NTT maps a polynomial from the coefficient domain to the transform domain, where polynomial multiplication becomes EWM~\cite{NTT,FFT}.
For a polynomial tower $x=(x_0,\ldots,x_{N-1})\in\mathbb{Z}_q^N$, the NTT can be written as $\hat{x}_j=\sum_{i=0}^{N-1}x_i\omega^{ij}\bmod q$, where $0\leq j<N$ and $\omega$ is an $N$-th primitive root of unity modulo $q$.
The INTT reconstructs the coefficient-domain polynomial as $x_i=N^{-1}\sum_{j=0}^{N-1}\hat{x}_j\omega^{-ij}\bmod q$.
Optimized CKKS implementations may use negacyclic variants, precomputed twiddle factors, and lazy reductions, but both NTT and INTT remain linear mappings over each RNS tower~\cite{RNS-CKKS,NTT}.

\textbf{BConv.}
BConv converts coefficients from one RNS basis to another~\cite{RNS-CKKS,CompositeScaling}.
Given an input basis $Q=\prod_iq_i$ and an output basis $P=\prod_jp_j$, the converted residue under modulus $p_j$ can be expressed as$y_j[k]=\sum_i x_i[k]\cdot\lambda_i\cdot\alpha_{i,j}\bmod p_j,$
where $x_i[k]$ is the $k$-th coefficient residue under $q_i$, and $\lambda_i$ and $\alpha_{i,j}$ are precomputed Chinese Remainder Theorem (CRT)-related constants.
BConv is coefficient-wise over polynomial positions, but it couples different RNS towers through the summation over $i$.

\textbf{EWM.}
After NTT, polynomial multiplication is reduced to EWM~\cite{CKKS,RNS-CKKS}.
Given two transformed polynomial towers $a=(a_0,\ldots,a_{N-1})$ and $b=(b_0,\ldots,b_{N-1})$, EWM computes $u_i=a_i\cdot b_i\bmod q$ for $0\leq i<N$.
Although EWM has linear complexity, it appears frequently in ciphertext multiplication, key switching, and bootstrapping~\cite{Bootstrapping,KLSS}.

\subsubsection{Ciphertext operators}

At the ciphertext level, CKKS supports addition, multiplication, rotation, and bootstrapping~\cite{CKKS,RNS-CKKS,Bootstrapping}.
Ciphertext addition is performed component-wise as $(c_0,c_1)+(c'_0,c'_1)=(c_0+c'_0,c_1+c'_1)\bmod Q$.
Ciphertext-plaintext multiplication multiplies each ciphertext component by an encoded plaintext polynomial.
Ciphertext-ciphertext multiplication first generates a three-component ciphertext, where $d_0=c_0c'_0\bmod Q$, $d_1=c_0c'_1+c_1c'_0\bmod Q$, and $d_2=c_1c'_1\bmod Q$.
The intermediate ciphertext $(d_0,d_1,d_2)$ is then relinearized back to two components through key switching~\cite{CKKS,RNS-CKKS,KLSS}.

Rotation applies an automorphism to permute encrypted slots, followed by key switching to restore the ciphertext under the original secret key.
Bootstrapping refreshes a ciphertext by homomorphically evaluating a decryption-like function, which invokes many rotations, key switches, NTT/INTT operations, BConv operations, and EWM kernels~\cite{Bootstrapping}.
Therefore, these high-level primitives ultimately rely on the same set of modular polynomial operators.

\subsubsection{Barrett modular reduction}

Modular reduction is a basic operation in CKKS arithmetic.
Since division by $q$ is expensive, CPU implementations commonly use Barrett reduction, which replaces division with multiplication and shift operations~\cite{barrett,ingrid_modular_reduction}.
For a modulus $q<2^w$ and an intermediate value $x$ with $0\leq x<2^{2w}$, Barrett reduction first precomputes$\mu=\left\lfloor \frac{2^{2w}}{q} \right\rfloor .$
Then, the quotient of $x/q$ is approximated by a multiplication and a shift, and the residue is obtained as
\begin{equation}
\label{eq:barrett_reduction}
t=\left\lfloor \frac{x\cdot \mu}{2^{2w}} \right\rfloor,\qquad
r=x-tq,\qquad
r\leftarrow \mathrm{Correct}(r,q),
\end{equation}
where $\mathrm{Correct}(r,q)$ denotes the final correction step that repeatedly subtracts or adds $q$ until $0\leq r<q$.
The resulting value satisfies $r\equiv x \pmod q$.

Although Barrett reduction avoids expensive integer division, invoking Eq.~\eqref{eq:barrett_reduction} repeatedly over long CKKS polynomials still causes significant overhead.
This motivates our later optimization that reduces the number of modular reductions during protection-side detection.

\subsection{Silent data corruptions}
\label{sec:pre_sdc}

Silent data corruptions (SDCs) refer to hardware-induced errors that produce incorrect outputs without triggering crashes, exceptions, or error reports~\cite{dixit2021silent,SDCExists,HardwareSDC}.
They may be caused by transient faults, marginal timing violations, manufacturing defects, or escaped design bugs~\cite{hochschild2021cores,dixit2021silent,parthasarathy2025sdc}.
Since no explicit failure signal is raised, corrupted values may silently propagate through the computation and only become observable at the final application output.

Existing protection techniques, such as ECC, parity checks, guardbanding, and redundant execution, can reduce hardware error rates, but they either mainly protect storage structures or introduce high overhead for computation-intensive workloads~\cite{udipi2012lotecc,zhang2018thundervolt}.
Algorithm-based fault tolerance (ABFT) provides a complementary approach by exploiting mathematical invariants of specific algorithms and checking the consistency between kernel inputs and outputs~\cite{ABFT}.

\textbf{Vulnerability of FHE computing.}
FHE computation is particularly sensitive to SDCs.
It executes long sequences of modular polynomial operations, which increases the fault-exposure window.
Meanwhile, ciphertexts are semantically opaque to the server, making intermediate corruptions difficult to detect.
A local bit flip in one coefficient or RNS residue may be further amplified by CRT reconstruction, NTT/INTT, basis conversion, and key switching~\cite{reliaanalysis,reliafhe}.
Therefore, efficient SDC detection is necessary for reliable FHE execution.

\textbf{Checksum for NTT.}
Checksum-based ABFT detects errors by verifying an algebraic invariant between the input and output of a kernel~\cite{ABFT,wang1994FFT,abdelmonem2025efficient}.
Since NTT is a linear transform, it naturally supports checksum protection~\cite{wang1994FFT,abdelmonem2025efficient}.
Let $\mathbf{y}=F\mathbf{x}$ denote an NTT over one RNS tower, where $F$ is the NTT transformation matrix.
The checker computes an input-side checksum $\mathrm{Enc}(\mathbf{x})=\sum_{i=0}^{N-1}e_ix_i\bmod q$ and an output-side checksum $\mathrm{Dec}(\mathbf{y})=\sum_{i=0}^{N-1}d_iy_i\bmod q$.
The encoding vector $\mathbf{e}$ and decoding vector $\mathbf{d}$ are selected to satisfy $\mathbf{e}^{T}=\mathbf{d}^{T}F$.
Therefore, under fault-free execution, $\mathrm{Enc}(\mathbf{x})=\mathbf{e}^{T}\mathbf{x}=\mathbf{d}^{T}F\mathbf{x}=\mathbf{d}^{T}\mathbf{y}=\mathrm{Dec}(\mathbf{y})$.
The NTT result is accepted only when $\mathrm{Enc}(\mathbf{x})\stackrel{?}{=}\mathrm{Dec}(\mathbf{y})$.

\subsection{Related work}
\label{sec:pre_related}

Existing fault studies on lattice-based cryptography mainly focus on security-oriented fault attacks, especially injecting faults during decryption to recover secret information~\cite{glitchfhe}.
Recent studies have also investigated the reliability of FHE ciphertext storage under memory faults~\cite{reliaanalysis}.
However, protecting FHE computation itself is different from protecting stored ciphertexts.
During homomorphic evaluation, errors can be generated and amplified by modular arithmetic, NTT/INTT, BConv, key switching, and bootstrapping.
Therefore, computation-level SDC detection requires operator-level and pipeline-level protection.

Recent FHE accelerator designs have explored efficient protection using customized datapaths, redundant modular units, or checksum-based invariants~\cite{reliafhe}.
These designs show that FHE arithmetic can be protected with lower overhead than full duplication.
However, their assumptions do not directly hold for CPU-based FHE libraries.
ASIC accelerators can introduce dedicated checksum datapaths and exploit massive customized parallelism, while CPUs must execute checksum logic through fixed instruction pipelines and cache hierarchies.
As a result, a checksum scheme that is efficient on an accelerator may still cause substantial instruction, modular-reduction, and memory-access overheads on CPUs.
This work therefore focuses on efficient checksum protection for CKKS software execution on CPUs.

\section{CKKS vulnerability and protection challenges on CPUs}
\label{sec:motivation_challenge}

This section motivates the need for efficient CPU-oriented error detection for CKKS computation.
We first evaluate the vulnerability of CKKS computations under transient faults.
We further show that, although existing redundancy and ABFT-based protection can improve reliability, directly applying them still incurs excessive overhead on CPUs.

\subsection{Vulnerability evaluation of CPU-based CKKS}
\label{sec:fault_model}


We first introduce the experimental setup for reliability evaluation on CPUs. Then, we evaluate the average accuracy of different FHE workloads under fault injection and analyze the error propagation model. Finally, we discuss the challenges of implementing efficient CKKS protection on CPUs.

\textbf{Evaluation setup.} To evaluate the vulnerability of FHE computation to transient faults, we select representative FHE workloads, inject faults at different stages of their execution, and measure the final application accuracy. 
For application-level evaluation, we consider four representative encrypted inference workloads, including LoLA, MLP, ResNet-20, and VGG16~\cite{LoLa,HELR,CKKSCNN,openfhe}.

We adopt a random single-bit transient fault model to evaluate the vulnerability of CPU-based FHE computation.
Using \texttt{PinFI}~\cite{wei2014pinfi}, faults are probabilistically injected into eligible dynamic instructions during ciphertext computation.
The fault rate denotes the probability that an eligible instruction is selected for fault injection.
Once selected, one randomly chosen bit in the instruction result is flipped.
To conform to the single-fault assumption of the operator-level protection scheme, each invocation of a low-level polynomial operator is restricted to at most one injected fault.
For each fault-rate setting, the experiment is repeated to characterize the impact of transient faults, and the resulting application accuracy is compared with that of the fault-free execution.

\begin{figure}[htbp]
    \centering
    \includegraphics[width=0.99\linewidth]{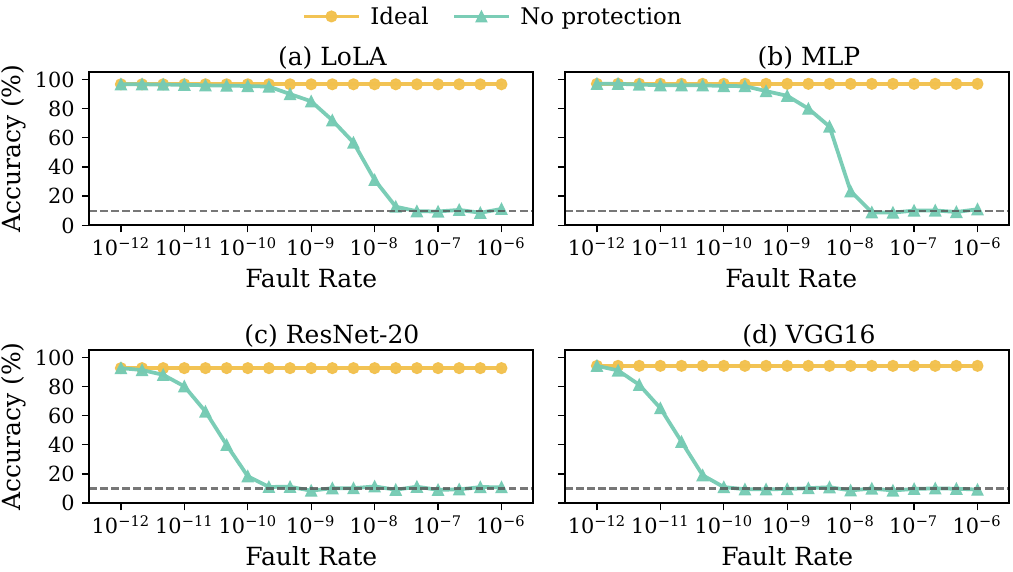}
    \caption{Vulnerability evaluation on four CKKS homomorphic applications.}
    \label{fig:app_vulnerability}
\end{figure}

\begin{figure}[htbp]
    \centering
    \includegraphics[width=0.7\linewidth]{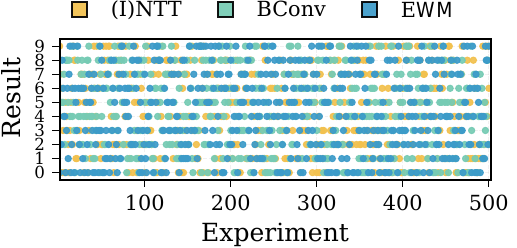}
    \caption{Vulnerability evaluation of fault injection on different polynomial operators.}
    \label{fig:poly_vulnerability}
\end{figure}

\textbf{Evaluation results.} Fig.~\ref{fig:app_vulnerability} shows the accuracy of four CKKS homomorphic applications under different fault rates.
The ideal fault-free execution maintains stable accuracy across all fault-rate settings.
In contrast, the unprotected CKKS executions degrade rapidly as the fault rate increases.
For LoLA and MLP, the accuracy starts to drop when the fault rate reaches around $10^{-9}$--$10^{-8}$.
For ResNet-20 and VGG-16, the degradation appears even earlier, and the final accuracy quickly approaches a nearly random level.
These results show that CKKS applications are highly sensitive to transient hardware faults.
Although CKKS is designed to tolerate cryptographic approximation noise, this tolerance does not naturally extend to arbitrary hardware-induced bit flips.

As described in Sec.~\ref{sec:preliminary}, FHE computation is composed of low-level polynomial operations. To further analyze how faults in different polynomial operators affect the final result, we inject faults into different polynomial operators during the execution of LoLA. 
As shown in Fig.~\ref{fig:poly_vulnerability}, faults in (I)NTT, BConv, and EWM operations can all lead to incorrect classification results. The vulnerable points are widely distributed across the polynomial pipeline rather than concentrated in a single operator. This indicates that protecting only one high-level ciphertext primitive or one selected polynomial operator is insufficient. An effective protection scheme should cover the major polynomial operations where errors are generated, transformed, and amplified.

\subsection{Challenge: protection overhead on CPUs}
\label{sec:challenge-cpu-overhead}

\begin{figure}[htbp]
    \centering
    \includegraphics[width=0.9\linewidth]{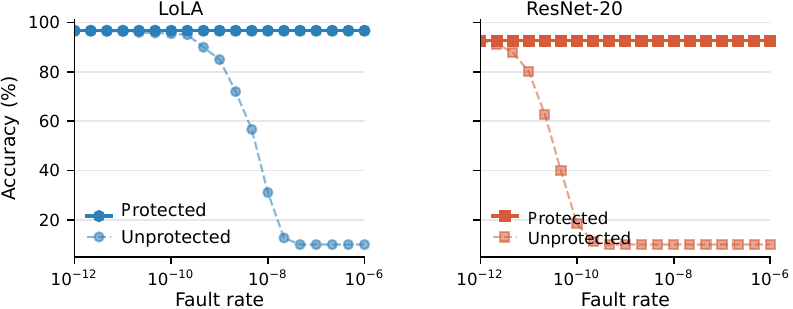}
    \caption{Reliability with basic protection under different fault rates.}
    \label{fig:abft_accuracy}
\end{figure}

Improving the reliability of FHE computation is feasible in principle.
A straightforward solution is redundant execution, which duplicates the original computation and compares the two results.
A more lightweight alternative is checksum-based error detection, which verifies the input--output consistency of polynomial operators using algorithmic invariants.
To evaluate the effectiveness of this approach on CPUs, we re-implement the checksum-based protection scheme proposed in~\cite{reliafhe} in our CPU-based FHE software and conduct fault-injection experiments.
As shown in Fig.~\ref{fig:abft_accuracy}, the basic checksum-based scheme substantially improves the reliability of encrypted inference.
For both LoLA and ResNet-20, the accuracy of unprotected execution rapidly degrades as the fault rate increases, whereas the protected execution maintains accuracy close to the fault-free baseline over a wide range of fault rates.

\begin{figure}[htbp]
    \centering
    \includegraphics[width=0.9\linewidth]{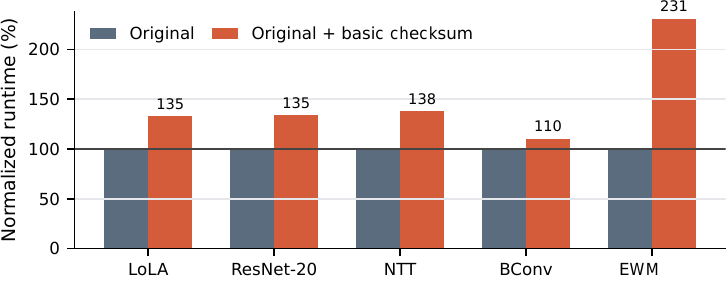}
    \caption{Runtime overhead of basic checksum-based protection on CKKS applications and operators.}
    \label{fig:abft_overhead}
\end{figure}

However, such protection is still too expensive for CPU-based FHE computing.
As shown in Fig.~\ref{fig:abft_overhead}, directly applying basic checksum-based protection introduces non-negligible overhead at both application and operator levels.
For end-to-end encrypted applications, the normalized runtime increases to $135\%$ for LoLA and ResNet-20.
For individual CKKS operators, the runtime increases to $138\%$ for NTT, $110\%$ for BConv, and $231\%$ for EWM.
Although this overhead is lower than that of full redundant
execution, which more than doubles the runtime in our evaluation,
it remains excessive for computationally intensive FHE workloads.
Therefore, the key challenge is not whether FHE computation can be protected, but how to achieve efficient error detection for FHE on CPUs.

\begin{figure}[htbp]
    \centering
    \includegraphics[width=0.99\linewidth]{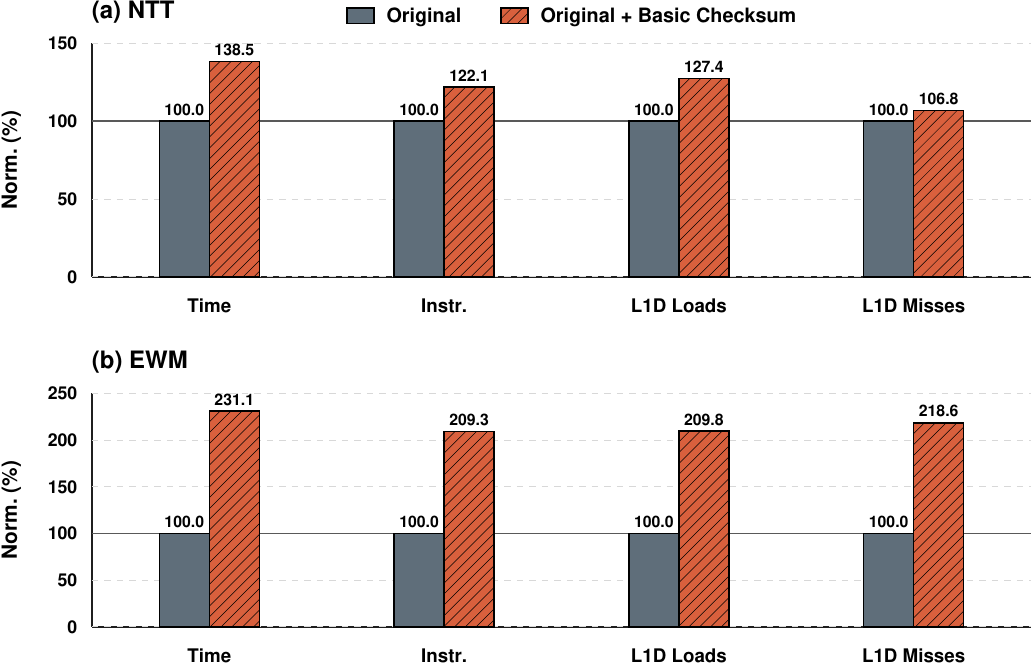}
    \caption{Hardware-counter profile of basic checksum-based protection for NTT and EWM.}
    \label{fig:abft_counter}
\end{figure}

To understand where this overhead comes from, we further profile the CPU execution behavior of basic checksum-based protection.
We report four normalized metrics in Fig.~\ref{fig:abft_counter}.
\emph{Time} measures the overall runtime overhead.
\emph{Instr.} reports the number of retired instructions, reflecting the extra arithmetic and control-flow work introduced by protection.
\emph{L1D Loads} measures the number of load accesses to the level-one data cache, reflecting the additional memory traffic.
\emph{L1D Misses} measures failed level-one data-cache accesses, indicating cache-locality degradation and increased pressure on the memory hierarchy.
All metrics are normalized to the original unprotected execution.

Fig.~\ref{fig:abft_counter} shows that basic checksum-based protection introduces both computation and memory overhead.
For NTT, the runtime increases to $138.5\%$, while the instruction count, L1D loads, and L1D misses increase to $122.1\%$, $127.4\%$, and $106.8\%$, respectively.
This indicates that NTT protection introduces moderate arithmetic overhead and additional memory accesses.
For EWM, the overhead is much more severe: the runtime increases to $231.1\%$, while the instruction count, L1D loads, and L1D misses increase to $209.3\%$, $209.8\%$, and $218.6\%$, respectively.
These results show that checksum execution on CPUs is constrained by both arithmetic cost and memory behavior.
Based on these results, we summarize three key observations that motivate our CPU-oriented protection design.

\textbf{Observation 1: checksum computation over modular polynomials incurs high arithmetic overhead.}
In ordinary integer domains, checksum-based checking mainly consists of simple multiply-accumulate operations.
However, CKKS operators process high-dimensional modular polynomials.
Each protection-side term may involve high-bit-width modular multiplication, Barrett modular reduction, and modular accumulation, rather than plain integer accumulation.
This introduces substantial instruction overhead on CPUs.
As shown in Fig.~\ref{fig:abft_counter}, the instruction count of protected NTT increases to $122.1\%$, and that of protected EWM further increases to $209.3\%$.
Therefore, reducing the modular arithmetic cost of protection-side computation is essential for efficient CPU-based FHE protection.

\textbf{Observation 2: checking long ciphertext polynomials introduces high memory-access overhead.}
CKKS ciphertexts contain long RNS polynomial towers with large coefficient bit-widths.
If checking is performed as a detached pass before or after the original operator, the protection logic needs to reload the operator inputs or outputs and traverse the long polynomial data again.
As shown in Fig.~\ref{fig:abft_counter}, protected NTT increases L1D loads to $127.4\%$, while protected EWM increases L1D loads and L1D misses to $209.8\%$ and $218.6\%$, respectively.
These results indicate that memory access is another major bottleneck of CPU-based protection.

\textbf{Observation 3: operator-by-operator checking introduces redundant verification in end-to-end FHE computation.}
Basic checksum-based protection usually protects each low-level polynomial operator independently.
This design is convenient because each operator can be checked through its own input-output relation, but it does not match the actual execution flow of FHE programs.
In CKKS, high-level primitives such as ciphertext multiplication, key switching, and bootstrapping are implemented as chains of polynomial operators, including INTT, BConv, NTT, and EWM.
If every operator independently computes and verifies protection values, the output check of one operator and the input check of the next operator may repeatedly process the same intermediate ciphertext data.
This causes redundant arithmetic and redundant memory accesses across adjacent operators.

In summary, accelerator-oriented protection schemes cannot be directly mapped to CPUs efficiently.
Their checking overhead can be mitigated by customized hardware datapaths and parallel execution in accelerators, but must be paid as extra instructions, memory accesses, and sequential software operations on CPUs.
Therefore, CPU-based FHE protection requires dedicated algorithmic and dataflow optimizations to reduce modular arithmetic, memory traffic, and redundant checks.
  \section{Proposed efficient protection scheme}
\label{sec:design}

\subsection{Overview} 
\label{sec:design_overview} 

\textbf{Complete protection mechanism.} 
This section presents our efficient fault-detection scheme for CPU-based CKKS computation.
We use polynomial operators as the basic detection granularity and protect each operator by checking the consistency between its inputs and outputs.
When adjacent operators have compatible checks, they are further grouped into short fused segments.
Once a mismatch is detected, only the affected operator or fused segment is re-executed, rather than the complete FHE task.

We first introduce the three optimization principles in Sec.~\ref{sec:checksum_optimization}. Sec.~\ref{sec:operator_protection} presents the protected NTT/INTT, EWM, and BConv operators. Sec.~\ref{sec:protected_pipeline} constructs the protected CKKS pipeline with cross-operator check fusion.

\subsection{Checksum optimizations for CPU-based CKKS computation}
\label{sec:checksum_optimization} 

To achieve efficient CKKS protection on CPUs, we propose three optimization methods:

\subsubsection{Methodology 1: modulus-aware bucket checksum}
\label{sec:modulus_aware_bucket}

\begin{figure}[htbp]
    \centering
    \includegraphics[width=0.99\linewidth]{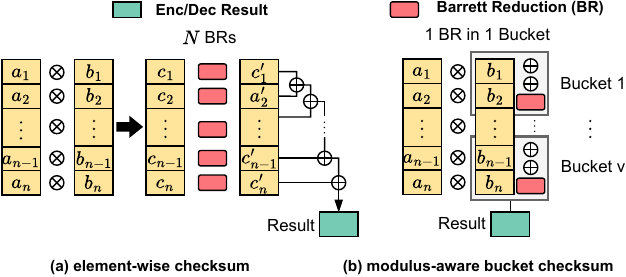}
    \caption{Modulus-aware bucket checksum.}
    \label{fig:bucket_checksum}
\end{figure}

As introduced in Sec.~\ref{sec:preliminary}, the basic checksum over a modular polynomial follows a coefficient-wise reduction flow.
As shown in Fig.~\ref{fig:bucket_checksum}(a), each product $a_i b_i$ is immediately reduced modulo $q$ before accumulation.
For an $N$-coefficient polynomial, this flow requires up to $N$ Barrett reductions, resulting in substantial arithmetic overhead on CPUs.

\textbf{Key idea.}
Our optimization is based on an important observation: the fixed-width wide intermediates provided by CPUs are not fully occupied by a single modular product in practical OpenFHE workloads.
We exploit the unused bit-width to accumulate multiple unreduced intermediate products, thereby constructing a bucketed checksum that performs one modular reduction for multiple coefficients.

\textbf{Bit-width utilization.}
In OpenFHE, each RNS coefficient is stored in a 64-bit word, while multiplying two coefficients already requires a 128-bit intermediate to hold the unreduced product.
Most RNS moduli in our evaluated CKKS configurations are approximately $50$--$60$ bits.
For $a_i,b_i\in[0,q-1]$, the product satisfies $a_i b_i \leq (q-1)^2$
and occupies at most $2\lceil\log_2 q\rceil$ bits.
Therefore, a single product occupies approximately $100$--$120$ bits, leaving about $8$--$28$ unused bits in the existing 128-bit intermediate.
As illustrated in Fig.~\ref{fig:bucket_checksum}(b), we use this remaining headroom to accumulate multiple products and invoke Barrett reduction only before another accumulation may cause overflow.
This optimization reuses the intermediate datatype already required by OpenFHE modular multiplication and introduces no wider arithmetic support.

\textbf{Modulus-aware pre-computation.}
Because each RNS tower has its own modulus $q$, its available accumulation headroom and safe bucket size are also different.
After the CKKS ciphertext parameters are determined, we pre-compute the bucket parameters separately for each modulus.
For an $N$-coefficient tower with modulus $q$ and a $D$-bit accumulator, the maximum safe bucket size $s(q)$ and the corresponding number of buckets $v(q)$ are defined as
\begingroup
\small
\begin{equation}
\label{eq:bucket_precomputation}
s(q)=
\left\lfloor
\frac{2^D-1}{(q-1)^2}
\right\rfloor,
\qquad
v(q)=
\left\lceil
\frac{N}{s(q)}
\right\rceil .
\end{equation}
\endgroup
Since each product is bounded by $(q-1)^2$, accumulating at most $s(q)$ products cannot overflow the $D$-bit accumulator.
In our implementation, $D=128$, matching the intermediate width already used by the original modular multiplication.
The values $s(q)$ and $v(q)$ are computed once during parameter initialization and reused for all subsequent operations on the corresponding RNS tower.
Therefore, no dynamic overflow check or repeated bucket-size computation is required during checksum evaluation.

\textbf{Bucketed checksum evaluation.}
Given a data vector $\mathbf{a}$ and a checksum vector $\mathbf{b}$ under modulus $q$, the proposed checksum is evaluated as
\begingroup
\small
\begin{equation}
\label{eq:bucket_checksum_algorithm}
C_q(\mathbf{a};\mathbf{b})
=
\left(
\sum_{j=0}^{v(q)-1}
\mathrm{BR}
\left(
\sum_{i=j s(q)}^{\min((j+1)s(q),N)-1}
a_i b_i,
q
\right)
\right)
\bmod q ,
\end{equation}
\endgroup
where $\mathrm{BR}(\cdot,q)$ denotes the Barrett reduction introduced in Eq.~\eqref{eq:barrett_reduction}.
Within each bucket, up to $s(q)$ products are accumulated without modular reduction.
The accumulated bucket sum is then reduced once modulo $q$.
Finally, the reduced results of all buckets are accumulated to obtain the final checksum.

\textbf{Correctness and benefit.}
The proposed computation preserves the original checksum because modular addition is linear.
For a partition of the coefficients into buckets $\{\mathcal{B}_j\}$,
\begingroup
\small
\begin{equation}
\label{eq:bucket_checksum_correctness}
\left(
\sum_{i=0}^{N-1} a_i b_i
\right)\bmod q
=
\left(
\sum_j
\left(
\sum_{i\in\mathcal{B}_j} a_i b_i
\right)\bmod q
\right)\bmod q .
\end{equation}
\endgroup
Thus, postponing modular reduction within each bucket does not change the checksum result or the original encode/decode invariant.

In summary, the proposed method converts the unused bit-width of existing CPU intermediates into useful accumulation capacity.
It reduces the number of checksum-side Barrett reductions from $N$ to $v(q)$, while requiring neither wider arithmetic types nor runtime overflow detection.
By pre-computing the bucket size according to each RNS modulus, the method adapts efficiently to different CKKS parameters with negligible runtime management overhead.

\subsubsection{Methodology 2: Dataflow-fused in-operator checking} \label{sec:dataflow_fused_check} 

\begin{figure}[htbp] 
\centering 
\includegraphics[width=0.99\linewidth]{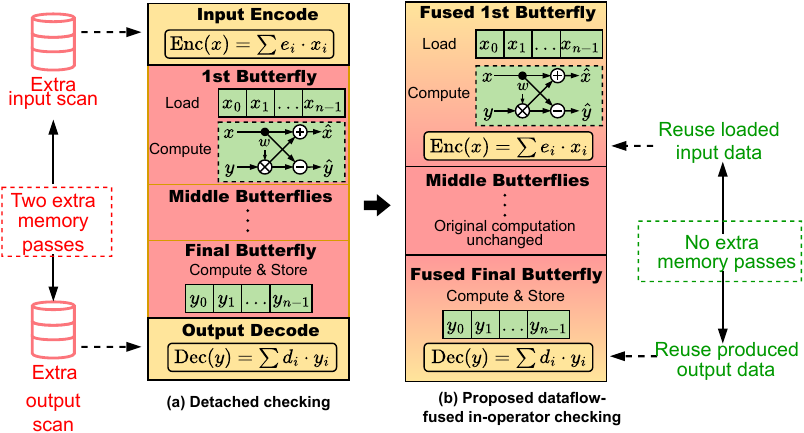} 
\caption{Dataflow-fused in-operator checking. Taking NTT as an example, the proposed design fuses input encoding and output decoding into the operator dataflow and reuses live data.} 
\label{fig:in_kernel_check} 
\end{figure} 

CKKS ciphertexts contain RNS polynomials with large sizes and high coefficient bit-widths. In OpenFHE, each coefficient is typically represented using a 64-bit word, and the polynomial degree is usually \(N=2^{14}\)--\(2^{16}\). As shown in Fig.~\ref{fig:in_kernel_check}(a), a detached checker scans the input polynomial for encoding and scans the output polynomial again for decoding, introducing extra memory passes over large polynomial data. This overhead has also been confirmed by the CPU profiling results in Sec.~\ref{sec:challenge-cpu-overhead}.

\textbf{Key idea.} To address this problem, we propose dataflow-fused in-operator checking, which moves encode/decode computation into the original operator dataflow. As illustrated in Fig.~\ref{fig:in_kernel_check}(b), the input-side encode is accumulated when coefficients are loaded by the operator implementation, and the output-side decode is accumulated when coefficients are produced and stored. Taking NTT as an example, the encode accumulation is fused into the first butterfly stage, while the decode accumulation is fused into the final butterfly stage. The middle butterfly stages remain unchanged because the checker only needs to relate the operator input and output. In this way, the checker reuses data that are already live in the operator dataflow, instead of launching detached input/output scans. 

\textbf{Correctness and benefit.} Dataflow-fused in-operator checking preserves the original encode/decode invariant; it only changes where the encode and decode terms are computed. By eliminating the two detached memory passes shown in Fig.~\ref{fig:in_kernel_check}(a), this optimization reduces memory traffic and improves cache locality.

\subsubsection{Methodology 3: Cross-operator check fusion}
\label{sec:cross_operator_check_fusion}

\begin{figure}[htbp]
    \centering
    \includegraphics[width=0.8\linewidth]{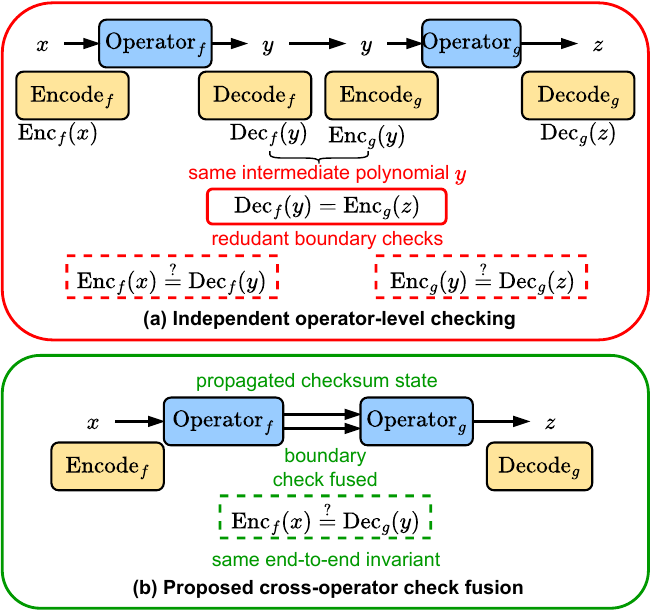}
    \caption{Cross-operator check fusion.}
    \label{fig:cross_operator_check}
\end{figure}

CKKS computation is a static pipeline composed of multiple polynomial operators. This static structure exposes opportunities to fuse redundant checks introduced by operator-by-operator protection. As shown in Fig.~\ref{fig:cross_operator_check}(a), for two adjacent operators \(f\) and \(g\), the baseline scheme performs a decode operation after \(f\) and another encode operation before \(g\) on the same intermediate polynomial \(\mathbf{y}\). If these two boundary checks use the same checksum vector, they repeatedly compute the same checksum value, causing redundant arithmetic and memory accesses.

\textbf{Key idea.}
We propose cross-operator check fusion to remove redundant boundary checks between adjacent operators.
Consider two consecutive operators $\mathbf{y}=f(\mathbf{x}), \mathbf{z}=g(\mathbf{y})$, the independent protection scheme verifies the two operators separately:
\begin{equation}
    \mathrm{Enc}_{f}(\mathbf{x})
    \stackrel{?}{=}
    \mathrm{Dec}_{f}(\mathbf{y}),
    \qquad
    \mathrm{Enc}_{g}(\mathbf{y})
    \stackrel{?}{=}
    \mathrm{Dec}_{g}(\mathbf{z}) .
    \label{eq:cross_op_independent}
\end{equation}
If the decode vector of \(f\) is identical to the encode vector of \(g\) for the same intermediate representation and RNS tower, then
\begin{equation}
    \mathrm{Dec}_{f}(\mathbf{y})
    =
    \mathrm{Enc}_{g}(\mathbf{y}) .
    \label{eq:cross_op_equiv}
\end{equation}
Therefore, the boundary decode/encode pair can be fused into one propagated checksum state.
As shown in Fig.~\ref{fig:cross_operator_check}(b), the fused pipeline removes the explicit \(\mathrm{Dec}_{f}(\mathbf{y})\) and \(\mathrm{Enc}_{g}(\mathbf{y})\) evaluations at the operator boundary.
Instead, it verifies the fused segment using the end-to-end relation:
\begin{equation}
    \mathrm{Enc}_{f}(\mathbf{x})
    \stackrel{?}{=}
    \mathrm{Dec}_{g}(\mathbf{z}) .
    \label{eq:cross_op_fused}
\end{equation}

\textbf{Correctness and benefit.}
Cross-operator fusion is applied only when the adjacent decode and encode operations share the same intermediate polynomial representation.
Under this condition, Eq.~\eqref{eq:cross_op_fused} is equivalent to composing the two independent checks in Eq.~\eqref{eq:cross_op_independent}, because the intermediate checksum state is identical by Eq.~\eqref{eq:cross_op_equiv}.
Thus, the fused segment preserves the same end-to-end encode/decode invariant and does not weaken error detection.
The benefit is that the redundant boundary check is removed.
When a mismatch is detected, the fused segment is re-executed from its last verified input.

\subsection{Operator-level protection for CKKS computation}
\label{sec:operator_protection}

In this section, we specify the protected invariant and the
corresponding protection algorithm for each operator:

\subsubsection{Protected NTT/INTT}
\label{sec:design_ntt}

For an $N$-coefficient polynomial under modulus $q$, let
$\mathbf{a}=(a_0,\ldots,a_{N-1})$
denote the input and
$\hat{\mathbf{a}}=\mathrm{NTT}(\mathbf{a})$
the transformed output.
According to the NTT checksum invariant introduced in Sec.~\ref{sec:preliminary}, the protected NTT verifies $C_{\mathrm{in}}
=
\sum_{i=0}^{N-1} a_i
\bmod q,
\qquad
C_{\mathrm{out}}
=
\sum_{i=0}^{N-1} e_i\hat a_i
\bmod q,$
where $\mathbf e=(e_0,\ldots,e_{N-1})$ is the decode vector determined by the NTT transform matrix and the implementation-specific output ordering.
The transform is accepted only if $C_{\mathrm{in}}
\stackrel{?}{=}
C_{\mathrm{out}}.$

To reduce the protection overhead, the NTT checker adopts the two optimizations introduced in Sec.~\ref{sec:checksum_optimization}: dataflow-fused in-operator checking and modulus-aware bucket checksum.

\begin{figure}[htbp]
\centering
\includegraphics[width=0.8\linewidth]{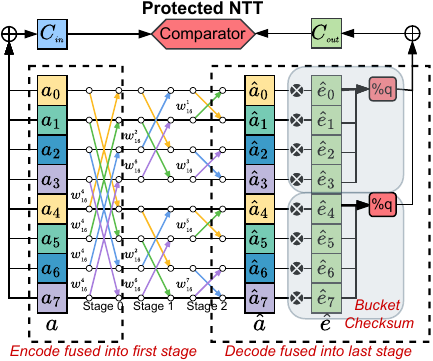}
\caption{Protected NTT dataflow with dataflow-fused in-operator checking and modulus-aware bucket checksum.}
\label{fig:protected_ntt}
\end{figure}

Algorithm~\ref{alg:protected_ntt} presents the protected implementation, while Fig.~\ref{fig:protected_ntt} illustrates how the checksum logic is integrated into the original NTT dataflow.
The butterfly computation itself remains unchanged; only checksum accumulation and comparison are inserted at the input and output boundaries.

\begin{algorithm}[t]
\small
\LinesNumbered
\DontPrintSemicolon

\KwIn{Input tower $a[0..N-1]$, modulus $q$, decode vector $e[0..N-1]$, bucket size $s(q)$, bucket number $v(q)$}
\KwOut{Output tower $\hat a=\mathrm{NTT}(a)$, fault flag}

\BlankLine

$C_{\mathrm{in}}\leftarrow0$ in a wide accumulator\;

\For{$b\leftarrow0$ \KwTo $v-1$}{
    $A_b\leftarrow0$ in a wide accumulator\;
}

\BlankLine

\For{each butterfly stage $r$}{

    \uIf{$r$ is the first stage}{

        \ForEach{input coefficient $a_i$ loaded by the NTT operator}{
            $C_{\mathrm{in}}\leftarrow C_{\mathrm{in}}+a_i$\;
        }

        execute the original first-stage butterflies\;
    }

    \uElseIf{$r$ is the final stage}{

        \ForEach{output coefficient $\hat a_i$ produced by the final butterfly}{

            store $\hat a_i$\;

            $b\leftarrow\lfloor i/s\rfloor$\;

            $A_b\leftarrow A_b+e_i\hat a_i$\;
        }

    }

    \Else{

        execute the original intermediate-stage butterflies\;

    }

}

\BlankLine

$C_{\mathrm{in}}\leftarrow C_{\mathrm{in}}\bmod q$\;

$C_{\mathrm{out}}
\leftarrow
\left(
\sum_{b=0}^{v-1}
\mathrm{BarrettReduce}(A_b,q)
\right)
\bmod q$\;

\If{$C_{\mathrm{in}}\neq C_{\mathrm{out}}$}{
    report fault\;
}
\Else{
    accept $\hat a$\;
}

\caption{Protected NTT with dataflow-fused in-operator checking and modulus-aware bucket checksum.}
\label{alg:protected_ntt}

\end{algorithm}

Lines~1--3 initialize the input checksum accumulator and the bucket accumulators.
These variables correspond to $C_{in}$ and the bucket checksum buffers shown at the top and right side of Fig.~\ref{fig:protected_ntt}.
Lines~5--8 correspond to the left part of Fig.~\ref{fig:protected_ntt}.
The input checksum is fused into the first butterfly stage:
whenever an input coefficient $a_i$ is loaded by the original NTT operator, it is simultaneously accumulated into $C_{\mathrm{in}}$.
Therefore, checksum computation reuses the original data-loading process without introducing an additional input scan.
Lines~9--13 correspond to the right part of Fig.~\ref{fig:protected_ntt}.
After the final butterfly produces $\hat a_i$, the checker immediately evaluates the weighted checksum term $e_i\hat a_i$ and accumulates it into the corresponding bucket $A_b$.
Each bucket performs unreduced multiply-accumulate operations and invokes Barrett reduction only once before overflow, following the modulus-aware bucket checksum described in Sec.~\ref{sec:modulus_aware_bucket}.
Finally, Lines~16--21 reduce all bucket sums to obtain $C_{\mathrm{out}}$ and compare it with $C_{\mathrm{in}}$.

The proposed implementation preserves the original NTT checksum invariant because dataflow fusion only changes the placement of checksum accumulation, while bucket checksum only postpones modular reduction from each coefficient to each bucket.
Consequently, detached checksum scans are eliminated and the number of checksum-side Barrett reductions is reduced from $N$ to $v(q)$.
The protected INTT is implemented in the same manner by replacing the NTT transform and decode vector with their inverse-transform counterparts.

\subsubsection{Protected EWM}
\label{sec:design_evalkey}

\begin{figure}[htbp] 
\centering 
\includegraphics[width=0.7\linewidth]{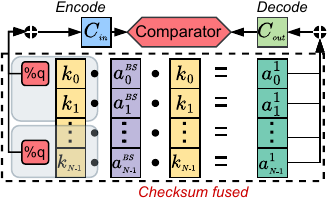} 
\caption{Protected EWM with dataflow-fused in-operator checking and modulus-aware bucket checksum.} 
\label{fig:protected_ewm} 
\end{figure} 

Unlike NTT, element-wise multiplication (EWM) does not naturally admit a low-cost linear checksum because every output coefficient depends on both input operands.
A straightforward checksum-based protection scheme therefore needs to independently evaluate the coefficient-wise products on the input side.
However, although a single EWM is cheaper than NTT, it is invoked frequently in ciphertext multiplication, key switching, and bootstrapping.
As shown in Sec.~\ref{sec:challenge-cpu-overhead}, directly protecting EWM introduces the largest runtime overhead among the evaluated polynomial operators.
Therefore, an efficient protection mechanism for EWM is essential.

For two input polynomials
$\mathbf a=(a_0,\ldots,a_{N-1})$,
$\mathbf k=(k_0,\ldots,k_{N-1})$,
and the output
$\mathbf a^{1}=(a_0^{1},\ldots,a_{N-1}^{1})$,
the protected EWM verifies $C_{\mathrm{in}}
=
\sum_{i=0}^{N-1}
a_i k_i
\bmod q,
\qquad
C_{\mathrm{out}}
=
\sum_{i=0}^{N-1}
a_i^{1}
\bmod q,$
and accepts the computation when
$
C_{\mathrm{in}}
\stackrel{?}{=}
C_{\mathrm{out}}.
$
Figure~\ref{fig:protected_ewm} illustrates the protected EWM dataflow.
Unlike NTT, EWM contains only one coefficient-wise multiplication stage.
Therefore, both checksum computation and operator execution naturally access the same coefficient stream, allowing checksum computation to be fully fused into the original operator.

During execution, every coefficient multiplication
$a_i k_i$
is immediately accumulated into the input checksum bucket, while the generated output coefficient
$a_i^{1}$
is simultaneously accumulated into
$C_{\mathrm{out}}$.
Following the modulus-aware bucket checksum introduced in Sec.~\ref{sec:modulus_aware_bucket}, the input checksum performs unreduced multiply-accumulate operations inside each bucket and invokes Barrett reduction only once per bucket.
The output checksum only requires direct accumulation because no coefficient multiplication is involved.

Compared with detached checksum evaluation, the proposed design eliminates additional memory scans and reduces the number of checksum-side Barrett reductions from $N$ to $v$.
Since the checksum computation is completely fused into the original coefficient-wise multiplication loop, the protection incurs only a small amount of additional arithmetic while preserving the original EWM checksum invariant.

\subsubsection{Protected BConv} 
\label{sec:design_bconv} 
Basis conversion maps input towers under \(Q=\{q_i\}\) to output towers under \(P=\{p_j\}\). For each output tower \(p_j\), we construct the encode result from the compressed input-tower checksums and construct the decode result from the actual output tower: $\mathrm{Enc}^{\mathrm{BConv}}_j = \sum_i \left( \sum_{k=0}^{N-1} X_i[k]\cdot \lambda_i \right) \alpha_{i,j} \pmod{p_j}, 
\mathrm{Dec}^{\mathrm{BConv}}_j = \sum_{k=0}^{N-1} Y_j[k] \pmod{p_j}, 
\mathrm{Enc}^{\mathrm{BConv}}_j \stackrel{?}{=} \mathrm{Dec}^{\mathrm{BConv}}_j, \qquad \forall p_j\in P .$

Here, \(X_i\) is the input tower under \(q_i\), \(Y_j\) is the output tower under \(p_j\), \(\lambda_i\) is the per-tower scaling constant used in basis conversion, and \(\alpha_{i,j}\) is the constant mapping from input tower \(q_i\) to output tower \(p_j\). This checksum is lightweight. On the decode side, the checker only accumulates the output coefficients of \(Y_j\). On the encode side, the checker first accumulates the coefficients of each input tower and then multiplies the compressed sum by the corresponding tower-level constants. Since the baseline BConv checksum is already much cheaper than NTT and EWM, this lightweight design is sufficient; after check fusion, the protection overhead of BConv is only about 3\%.

For other lightweight operators, such as ciphertext addition and polynomial
automorphism, we adopt low-cost redundancy-based protection with check fusion.
These operators have relatively low computational complexity and account
for only a small fraction of the end-to-end CKKS runtime.
Therefore, it introduces limited overall overhead.

\subsection{Protected CKKS pipeline with cross-operator check fusion}
\label{sec:protected_pipeline}

\begin{figure}[htbp]
    \centering
    \includegraphics[width=0.9\linewidth]{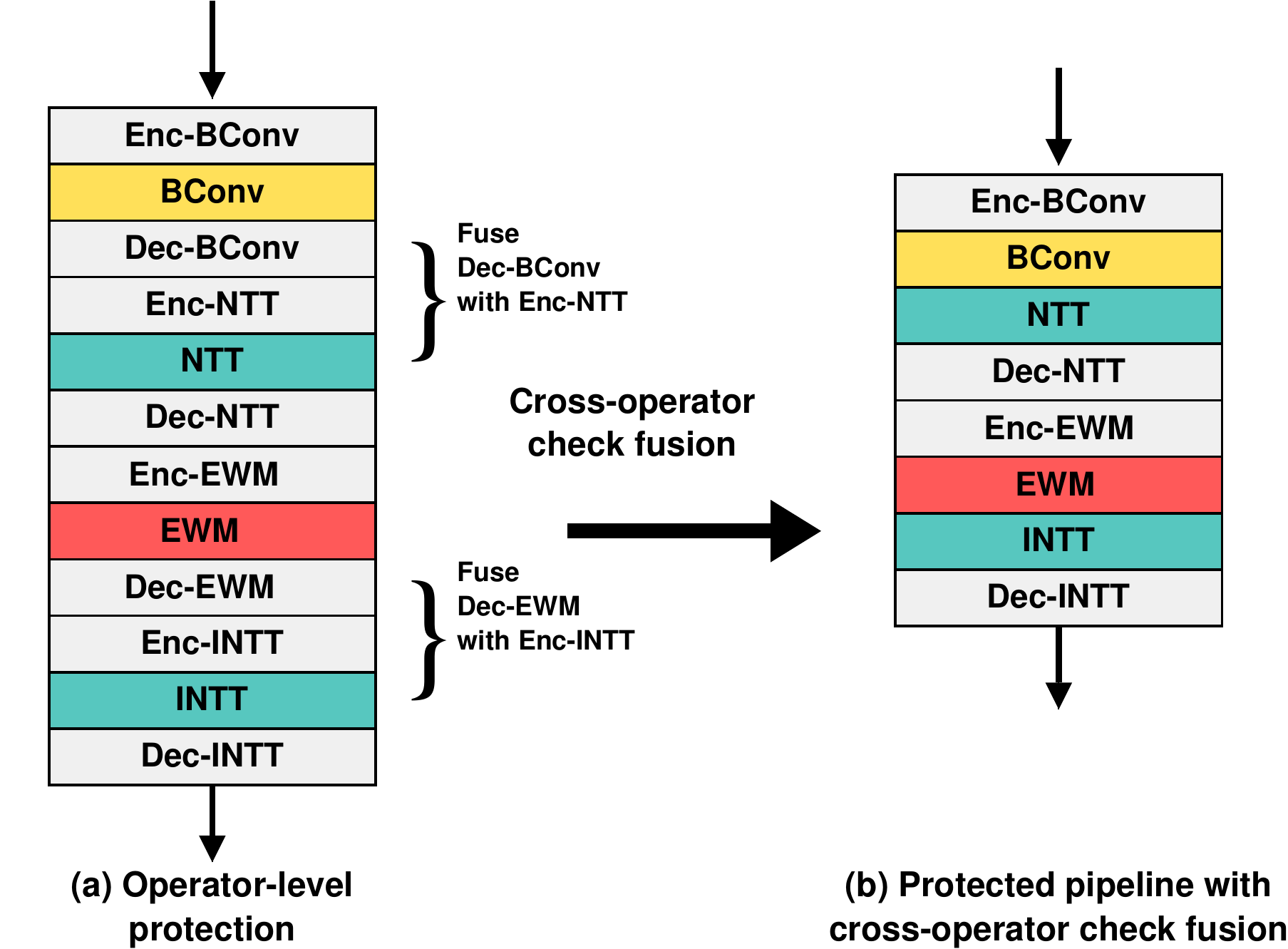}
    \caption{Protected CKKS pipeline with cross-operator check fusion.}
    \label{fig:cross_op_pipeline}
\end{figure}

\textbf{Pipeline construction.}
The operator-level protection schemes described above can be connected to construct a fully protected CKKS computation pipeline.
Fig.~\ref{fig:cross_op_pipeline}(a) shows a representative segment of the core polynomial-operator pipeline in CKKS ciphertext multiplication, where BConv, NTT, EWM are executed sequentially.
Under the operator-level protection, each operator is independently surrounded by an input-side encode and an output-side decode.
Although this design provides fine-grained error detection, the pipeline structure exposes opportunities for cross-operator check fusion: the output decode of one operator and the input encode of its successor may compute the same checksum over the shared intermediate polynomial.

\textbf{Fusion instances.}
According to the cross-operator fusion rule introduced in
Sec.~\ref{sec:cross_operator_check_fusion}, for two adjacent operators
$\mathbf{y}=f(\mathbf{x})$ and $\mathbf{z}=g(\mathbf{y})$, if $\mathrm{Dec}_{f}(\mathbf{y})
=
\mathrm{Enc}_{g}(\mathbf{y})$, the intermediate decode and encode computations can be removed.
The two independent checks are then replaced by a composed check from
$\mathrm{Enc}_{f}(\mathbf{x})$ to $\mathrm{Dec}_{g}(\mathbf{z})$, as
formulated in
Eqs.~\eqref{eq:cross_op_independent}--\eqref{eq:cross_op_fused}.

In the CKKS pipeline shown in Fig.~\ref{fig:cross_op_pipeline}(a), this
condition holds at two operator boundaries.
First, the output decode of BConv and the input encode of NTT are both
coefficient-sum checks over the same RNS tower.
Second, the output decode of EWM and the input encode of INTT use the same
coefficient-sum relation: $\mathrm{Dec}^{(t)}_{\mathrm{BConv}}
\bigl(\mathbf{y}^{(t)}\bigr)=
\sum_{i=0}^{N-1} y_i^{(t)}
\bmod q_t
=
\mathrm{Enc}^{(t)}_{\mathrm{NTT}}
\bigl(\mathbf{y}^{(t)}\bigr),
\mathrm{Dec}^{(t)}_{\mathrm{EWM}}
\bigl(\mathbf{u}^{(t)}\bigr)=
\sum_{i=0}^{N-1} u_i^{(t)}
\bmod q_t
=
\mathrm{Enc}^{(t)}_{\mathrm{INTT}}
\bigl(\mathbf{u}^{(t)}\bigr).$

Accordingly, the optimized pipeline in
Fig.~\ref{fig:cross_op_pipeline}(b) removes the redundant
$\mathrm{Dec}_{\mathrm{BConv}}$--$\mathrm{Enc}_{\mathrm{NTT}}$ and
$\mathrm{Dec}_{\mathrm{EWM}}$--$\mathrm{Enc}_{\mathrm{INTT}}$
boundary computations.
The intermediate comparisons are eliminated, while the composed checksum
relations are verified at the outputs of the corresponding fused segments.
Other operator boundaries remain independently checked when their decode
and encode relations are incompatible.

\textbf{Analysis.}
Cross-operator fusion composes compatible operator-level invariants without
changing the end-to-end checksum relation.
Thus, fusion postpones the comparison from the intermediate operator
boundary to the end of the fused segment, rather than removing error
detection.
This optimization introduces a trade-off in fault-localization and recovery granularity.
With independent operator-level checking, a mismatch requires re-execution of only the affected operator.
After fusion, the recovery scope expands to a short segment containing two adjacent polynomial operators.
Nevertheless, detection and recovery remain at the polynomial-operator level, and the cost of re-executing is still small compared with the runtime of a complete FHE computation.

Taken together, these three techniques reduce checksum overhead
at the arithmetic, operator-dataflow, and operator-pipeline levels,
enabling efficient protection for FHE programs running on CPUs.


\section{Experimental evaluation}
\label{sec:evaluation}

We first present the experimental setup in Section~\ref{sec:eval_set}, including the evaluation platform, workloads, and fault models. Then, Section~\ref{sec:eval_reliability} provides the reliability evaluation. Finally, Section~\ref{sec:eval_performance} presents the performance evaluation and ablation study.

\subsection{Evaluation setup}
\label{sec:eval_set}

\textbf{FHE benchmarks and platform.}
We evaluate the proposed protection scheme at both application and primitive
levels.
For end-to-end evaluation, we select four representative CKKS encrypted
inference workloads: LoLA, MLP, ResNet-20, and VGG-16.
For primitive-level evaluation, we consider ciphertext--ciphertext
multiplication (HMult) under two parameter configurations and CKKS
bootstrapping.
The parameters used for the primitive-level experiments are summarized in
Table~\ref{tab:primitive_params}.
All experiments are conducted on an Intel(R) Xeon(R) Platinum 8358P CPU. 
We use the official  OpenFHE~\cite{openfhe} and compile all experiments in Release mode. 

\begin{table}[htbp]
\centering
\caption{CKKS parameters used for primitive-level evaluation.
$L$ denotes the multiplicative depth, $Q$ denotes the ciphertext modulus,
and $\lambda$ denotes the security level.}
\label{tab:primitive_params}
\scriptsize
\setlength{\tabcolsep}{6pt}
\renewcommand{\arraystretch}{1.12}
\begin{tabular*}{\columnwidth}{
@{\extracolsep{\fill}}lccccc}
\hline\hline
\textbf{Configuration}
& $N$
& \textbf{Slots}
& $L$
& $\log_2 Q$
& $\lambda$ \\
\hline
HMult Set 1
& $2^{16}$
& $2^{15}$
& $30$
& $\approx 1560$
& $128$ \\

HMult Set 2
& $2^{15}$
& $2^{14}$
& $20$
& $\approx 700$
& $128$ \\

Bootstrapping
& $2^{16}$
& $2^{15}$
& $30$
& $\approx 1560$
& $128$ \\
\hline\hline
\end{tabular*}
\end{table}

\textbf{Fault model and reliability evaluation.}
We evaluate both encrypted applications and ciphertext primitives under the
random single-bit transient fault model described in
Sec.~\ref{sec:fault_model}.
Using \texttt{PinFI}~\cite{wei2014pinfi}, faults are probabilistically injected into eligible
dynamic instructions during ciphertext computation.
The fault rate denotes the probability that an eligible instruction is
selected, after which one random bit in its result is flipped.
We restrict each invocation of a low-level polynomial operator to at most one
injected fault.



\textbf{Performance overhead measurement.}
We evaluate runtime overhead on the encrypted applications and ciphertext
primitives described above.
For each workload, the unprotected and protected implementations are executed
under the same configuration for 1,000 runs, and the average runtime is
reported.
We compare our method with dual modular redundancy (DMR) and a basic
checksum-based baseline.
DMR executes each polynomial operation twice and compares the two outputs.
Since ReliaFHE~\cite{reliafhe} is designed for ASIC accelerators and
is not open-sourced, we reproduce its algorithmic checksum scheme in
OpenFHE without its customized hardware support.

\subsection{Reliability evaluation}
\label{sec:eval_reliability}

\begin{figure}[htbp]
    \centering
    \includegraphics[width=0.99\linewidth]{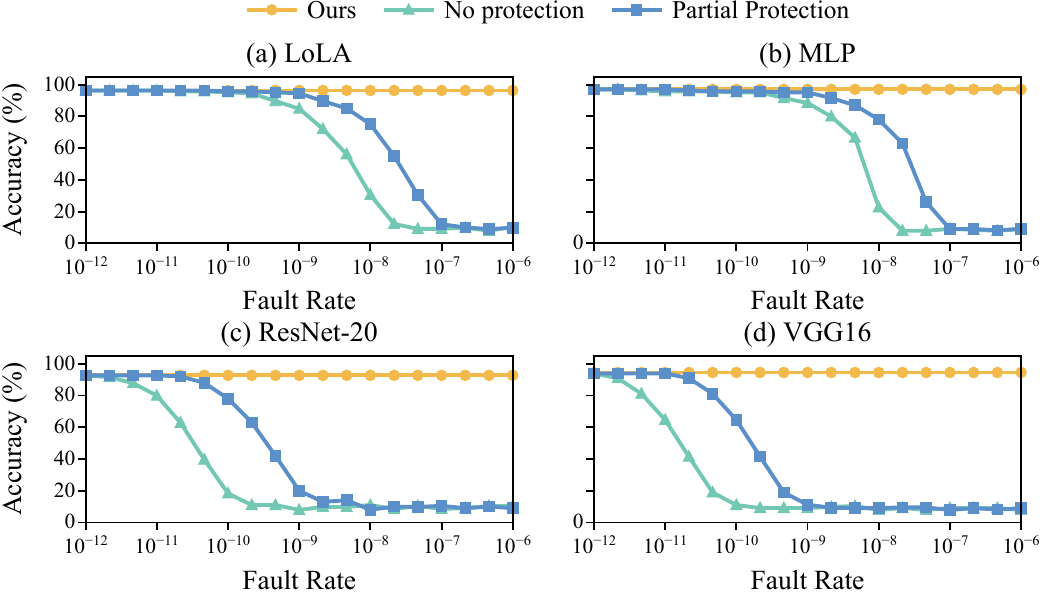}
    \caption{Application-level reliability under different fault rates.}
    \label{fig:reliab_eval}
\end{figure}

We evaluate the reliability of the proposed scheme from two perspectives.
First, we measure the end-to-end accuracy of encrypted applications under
different fault rates.
Second, we evaluate the SDC detection rate for faults occurring in the major
polynomial operations.

\begin{figure*}[b]
    \centering
    \includegraphics[width=0.99\linewidth]{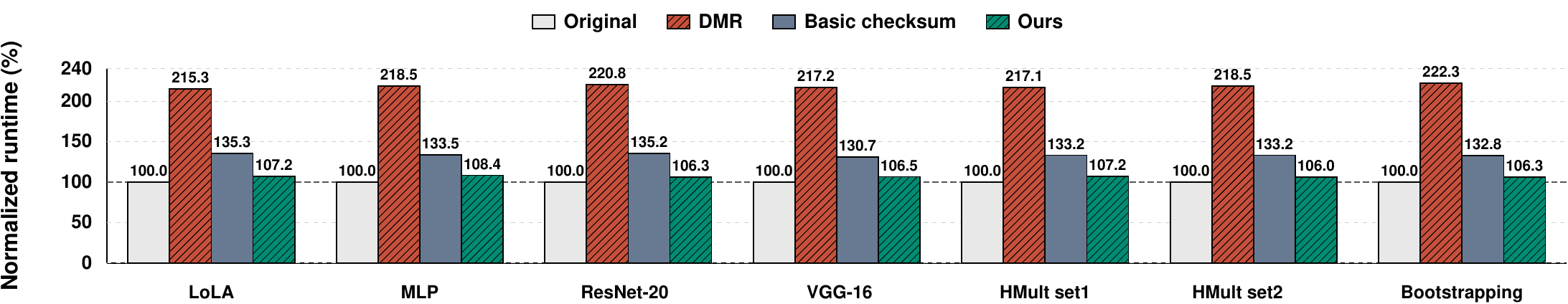}
    \caption{Overall performance comparison}
    \label{fig:perf_eval}
\end{figure*}

\textbf{Application-level reliability.}
Fig.~\ref{fig:reliab_eval} compares three configurations on LoLA, MLP,
ResNet-20, and VGG-16: no protection, partial protection that protects only
NTT/INTT and BConv, and the proposed full-protection scheme.
When a mismatch is detected, the corrupted intermediate result is discarded
and the corresponding polynomial operator or fused segment is re-executed.

A consistent trend can be observed across all four applications.
Without protection, the application accuracy decreases rapidly as the fault
rate increases, demonstrating that transient faults in low-level polynomial
computation can propagate through the ciphertext dataflow and severely
corrupt the final inference result.
The degradation occurs particularly early for the deeper ResNet-20 and
VGG-16 workloads, which contain longer ciphertext-computation flows and
therefore expose more opportunities for error propagation.

Partial protection consistently delays the accuracy degradation compared
with the unprotected execution.
However, its accuracy still decreases substantially at higher fault rates
because faults occurring in unprotected polynomial operations can escape
detection.
Therefore, protecting only selected operators improves reliability but
cannot ensure dependable end-to-end execution.

In contrast, the proposed full-protection scheme maintains nearly unchanged
accuracy throughout the evaluated fault-rate range for all four applications.
The clear separation between our method and the other two configurations
shows that comprehensive protection of the major polynomial operations is
necessary to prevent fault-induced errors from propagating to the final FHE
output.


\textbf{Error detection rate.}
We further evaluate the error-detection capability of the proposed
operator-level checks using random single-fault injection.
For each operator category, one random single-bit transient fault is
injected into each evaluated operator invocation.
Trials that directly cause program crashes are excluded.
We continue the experiments until 50,000 non-crashing error cases are
collected for each operator category.

An error is considered detected when the injected fault corrupts the
operator result and triggers a checksum mismatch before the corrupted result
is consumed by subsequent computation.
As shown in Table~\ref{tab:error_detection}, the proposed scheme achieves a 100\% empirical detection rate over 150,000 collected non-crashing corrupted-result cases under the evaluated single-fault model.
The reported 100\% detection rate is empirical and applies to
the collected non-crashing cases under the
evaluated fault model. As with
 checksum-based detection, specially structured multiple
errors that preserve the checksum relation may escape detection
and are outside the fault model considered in this work.

Together with the application-level results, these experiments demonstrate
that the proposed protection scheme effectively detects errors in the major
CKKS polynomial operations and prevents corrupted intermediate results from
silently propagating to the final application output.

\begin{table}[t]
\centering
\caption{Error detection rate under random single-fault injection.}
\label{tab:error_detection}
\scriptsize
\setlength{\tabcolsep}{5pt}
\renewcommand{\arraystretch}{1.12}
\begin{tabular*}{\columnwidth}{
@{\extracolsep{\fill}}lccc}
\hline\hline
\textbf{Fault location}
& \textbf{Error cases}
& \textbf{Detected errors}
& \textbf{Detection rate} \\
\hline
NTT/INTT
& 50,000
& 50,000
& $100\%$ \\

BConv
& 50,000
& 50,000
& $100\%$ \\

EWM
& 50,000
& 50,000
& $100\%$ \\
\hline
\textbf{Total}
& \textbf{150,000}
& \textbf{150,000}
& \textbf{100\%} \\
\hline\hline
\end{tabular*}
\end{table}

\subsection{Performance evaluation} \label{sec:eval_performance}

We evaluate the steady-state runtime overhead of the proposed protection
scheme without fault injection.
The benchmarks include four encrypted applications, two HMult
configurations, and CKKS bootstrapping, as specified in
Table~\ref{tab:primitive_params}.
We first compare the end-to-end overhead with dual modular redundancy (DMR)
and basic checksum protection, then perform a progressive ablation study,
and finally analyze the operator-level execution behavior using hardware
counters.

\begin{figure*}[t]
    \centering
    \includegraphics[width=0.99\linewidth]{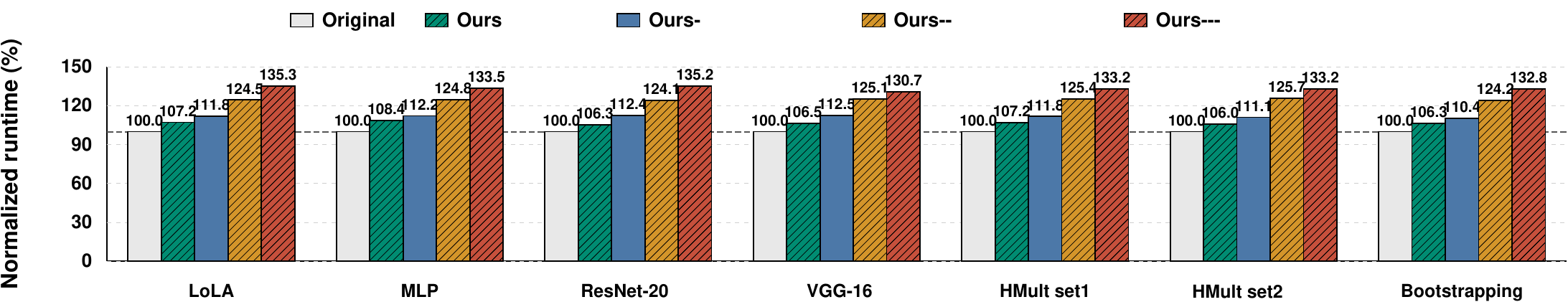}
    \caption{Progressive ablation study on protection overhead.
    \emph{Ours} denotes the complete design;
    \emph{Ours-} removes cross-operator check fusion;
    \emph{Ours-{}-} further removes modulus-aware bucket checksum;
    and \emph{Ours-{}-{}-} additionally removes dataflow-fused in-operator checking,
    thereby degenerating into the basic detached checksum implementation.}
    \label{fig:ablation_overhead}
\end{figure*}

\begin{figure*}[t]
    \centering
    \includegraphics[width=0.9\linewidth]{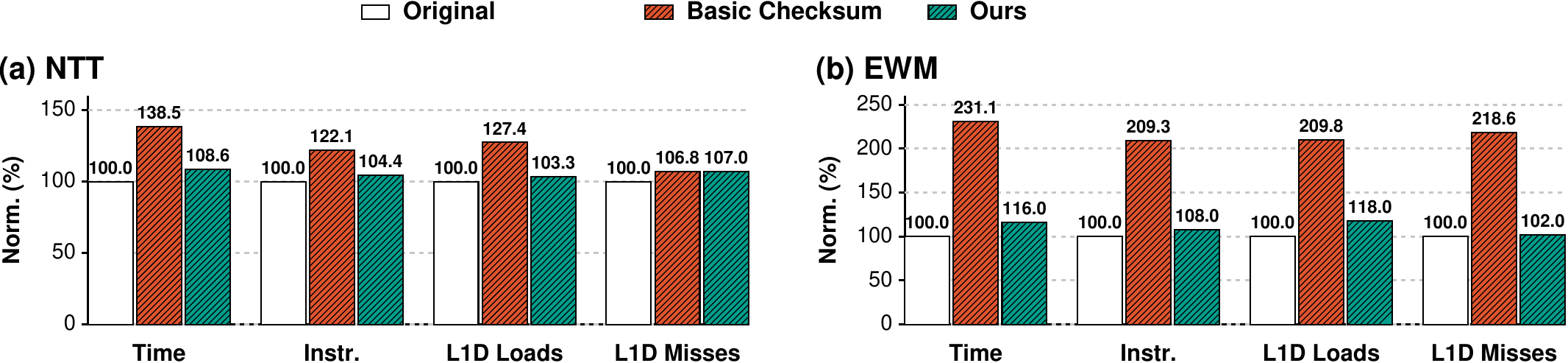}
    \caption{Fine-grained hardware-counter profile of checksum protection on NTT and EWM.}
    \label{fig:fine_grained_counter}
\end{figure*}

\textbf{Overall performance comparison.}
Fig.~\ref{fig:perf_eval} compares the normalized runtime of the different
protection schemes, where the original unprotected execution is normalized
to 100\%.
The results show a consistent trend across all encrypted applications and
ciphertext primitives.

DMR incurs the highest overhead, increasing the normalized runtime to
\(215.3\%\)--\(222.3\%\).
This is expected because DMR duplicates each protected computation and
additionally compares the two results, resulting in more than twice the
original runtime.

Basic checksum protection is substantially more efficient than DMR, but
still increases the normalized runtime to
\(130.7\%\)--\(135.3\%\).
Although it avoids full computation duplication, its detached checksum
evaluation introduces additional modular arithmetic and repeated scans over
large RNS polynomial towers.
Therefore, its algorithmic advantage does not directly translate into low
execution overhead on CPUs.

In contrast, our method consistently limits the normalized runtime to
\(106.0\%\)--\(108.4\%\) across all seven benchmarks.
On average, the proposed scheme incurs only \(6.8\%\) additional runtime,
compared with \(33.4\%\) for basic checksum protection.
This corresponds to a \(4.9\times\) reduction in average protection
overhead.
The consistently low overhead across applications, HMult configurations,
and bootstrapping demonstrates that the proposed CPU-oriented optimizations
remain effective across different CKKS workloads and parameter settings.

\textbf{Progressive ablation study.}
Fig.~\ref{fig:ablation_overhead} evaluates the contribution of the three
proposed optimizations through progressive ablation.
Starting from the complete design (\emph{Ours}), we cumulatively remove
cross-operator check fusion (\emph{Ours-}), modulus-aware bucket checksum
(\emph{Ours-{}-}), and dataflow-fused in-operator checking (\emph{Ours-{}-{}-}).
The original unprotected execution is normalized to 100\%.

The complete design (\emph{Ours}) consistently achieves low overhead.
After removing cross-operator check fusion (\emph{Ours-}), the normalized
runtime increases to \(110.4\%\)--\(112.5\%\).
Without cross-operator fusion, the output-side check of one operator and the
input-side check of its successor must process the same intermediate
polynomial separately, leading to additional checksum computation.

When modulus-aware bucket checksum is further removed (\emph{Ours-{}-}), the
normalized runtime increases more substantially to
\(124.1\%\)--\(125.7\%\).
In this configuration, checksum products are reduced coefficient by
coefficient instead of being accumulated in wide buckets and reduced once
per bucket.
The pronounced increase confirms that frequent modular reduction is a major
source of protection overhead on CPUs, and validates the effectiveness of
the proposed bucket checksum in reducing checksum-side modular arithmetic.

Finally, after dataflow-fused in-operator checking is also removed
(\emph{Ours-{}-{}-}), the implementation degenerates into the basic detached
checksum baseline.
The normalized runtime further increases to \(130.7\%\)--\(135.3\%\).
This additional overhead comes from detached checksum computation outside
the original operator dataflow, which introduces standalone scans over
large RNS polynomial and reduces data locality.
The result shows that in-operator fusion is important for avoiding extra
memory accesses and improving cache reuse.
Overall, the progressive ablation study demonstrates that all three
optimizations are necessary for the final low-overhead design.

\textbf{Further analysis on operator-level overhead.}
To connect the end-to-end results with the CPU bottlenecks identified in
Sec.~\ref{sec:challenge-cpu-overhead}, we profile the representative NTT
and EWM operators using hardware counters.
As shown in Fig.~\ref{fig:fine_grained_counter}, our design reduces the
runtime of protected NTT from \(138.5\%\) to about \(108.6\%\), while bringing
the instruction count and L1D loads close to the unprotected execution.
For EWM, the runtime decreases from \(231.1\%\) to \(116.0\%\), together
with substantial reductions in instructions, L1D loads, and L1D misses.
These results confirm that modulus-aware bucket checksum reduces modular
arithmetic overhead, while in-operator fusion eliminates most detached
polynomial scans.
The protection overhead of BConv is about \(3\%\).
Together, these operator-level improvements reduce the end-to-end normalized
runtime of the protected applications to approximately
\(106\%\)--\(108\%\).

\textbf{Fault-recovery overhead.}
When an error is detected, only the affected polynomial operator or short
fused segment is re-executed.
Across the evaluated workloads, one recovery accounts for less than 1.0\% of the end-to-end runtime.
  \section{Limitations and Future Work}
\label{sec:limitations}

This work has several limitations.
First, our evaluation focuses on random single-bit transient faults and
assumes at most one injected fault within each polynomial-operator
invocation.
Although this model covers common transient computation
errors, it does not fully represent multiple or correlated faults
within the same operator invocation.
Future work will extend the protection framework and fault-injection
evaluation to these more complex fault models.

Second, the current implementation targets CKKS computation in OpenFHE on
one CPU platform.
Although the proposed optimizations exploit general properties of
RNS-based polynomial computation, their effectiveness on other FHE schemes,
software libraries, CPU architectures, and parallel execution environments
requires further evaluation.
Future work will also investigate automatic invariant construction and
adaptive protection granularity for broader FHE workloads.

\section{Conclusion}
\label{sec:conclusion}

This paper presents an efficient fault-tolerance scheme for CKKS
computation on general-purpose CPUs.
To reduce the arithmetic, memory-access, and redundant-checking
overheads of direct checksum protection, we propose modulus-aware
bucket checksum, dataflow-fused in-operator checking, and
cross-operator check fusion.
Together, these techniques reduce modular reductions, eliminate
detached polynomial scans, and remove redundant checks between
adjacent operators.

We implement the proposed scheme in OpenFHE and evaluate it on
representative encrypted applications and CKKS primitives.
Under the evaluated random single-bit transient fault model, the
scheme detects all collected non-crashing corrupted-result cases
and maintains application accuracy close to the fault-free baseline.
It incurs only \(6.0\%\)--\(8.4\%\) runtime overhead, averaging
\(6.8\%\), and reduces the average protection overhead by
approximately \(4.9\times\) compared with basic checksum
protection.
These results demonstrate that targeted checksum and dataflow
optimizations can enable efficient and reliable CKKS computation
on general-purpose CPUs.

  \bibliographystyle{IEEEtran}
  \bibliography{ref/ref}



  \end{document}